\documentclass[cameraready]{Interspeech}

\title{SE-AGCNet: An End-to-End Framework for Joint Speech Enhancement and Loudness Control in Meeting Scenarios}

\author[affiliation={1}]{Jinming}{Zhang}
\author[affiliation={2}]{Wei}{Rao}
\author[affiliation={3}]{Xionghu}{Zhong}
\author[affiliation={2}]{Eng Siong}{Chng}

\address{
    $^1$ Zhejiang University, China \\
    $^2$ Nanyang Technological University, Singapore \\
    $^3$ Hunan University, China
}

\email{pmhuan1212@gmail.com, aseschng@ntu.edu.sg}

\usepackage{comment}
\usepackage[utf8]{inputenc}
\usepackage{graphicx}  
\usepackage{booktabs}  
\usepackage{multirow}  
\usepackage{makecell}   
\usepackage{siunitx}    
\usepackage{adjustbox} 
\usepackage{arydshln}  

\begin{document}

\maketitle

%

\begin{abstract}
Conventional audio pipelines typically treat speech enhancement (SE) and automatic gain control (AGC) as discrete modules, which often limits overall performance. For instance, applying AGC before SE may inadvertently amplify background noise, while prioritizing SE tends to over-suppress low-volume speech. To address these limitations, we propose SE-AGCNet, an end-to-end framework that jointly optimizes SE and AGC. Tailored for meeting scenarios with significant volume variations, SE-AGCNet leverages the synergy between the two tasks: SE preserves quiet speech, thereby facilitating effective volume adjustment by the AGC component. Furthermore, we propose a specialized data simulation pipeline, SE-AGC-DataGen, and incorporate standardized loudness evaluation metrics: integrated loudness (LUFS), short-term loudness (St LUFS), and LRA. Experiments show that SE-AGCNet consistently achieves target loudness while improving speech quality and ASR accuracy over competitive baselines.
\end{abstract}


\section{Introduction}
\label{sec:intro}

The audio front-end is traditionally characterized by the "3A algorithms": Acoustic Echo Cancellation (AEC), Noise Suppression (also referred to as speech enhancement), and Automatic Gain Control (AGC). Conventional audio pipelines typically implement AGC~\cite{wang2022automatic,garcia2020automatic,ambeth2020active,iwai2019audio,yang2018multilayer,yang2017deep,sugiyama2017automatic,petkov2017adaptive,nagata2005speech,archibald2008software,prabhavalkar2015automatic,sredojev2015webrtc} and Speech Enhancement (SE) \cite{lu2025explicit,yin23_interspeech,yao2025gense} as separate, cascaded modules, a design that introduces several limitations. When AGC precedes SE, it amplifies speech and noise indiscriminately, thereby reducing the signal-to-noise ratio and complicating subsequent denoising. When applied after SE, AGC performance becomes highly dependent on SE quality, as residual noise is often unintentionally amplified. A further challenge arises from SE models' tendency to over-suppress low-volume speech.

Jointly training SE and AGC provides a principled solution to the limitations of cascaded pipelines. Through simultaneous optimization, the SE module is encouraged to preserve low-volume speech while concentrating on noise suppression, with the assurance that volume normalization will be managed by the AGC module. This synergistic framework mitigates two key issues: it reduces the risk of SE misclassifying far-field or quiet speech as noise, and it prevents AGC from inadvertently amplifying residual background noise.

Recent studies have attempted to integrate AEC, SE, and AGC within unified frameworks. For example,~\cite{wang2022nn3a} introduced a unified 3A framework but treated AGC merely as a detached post-processing module without specifying its underlying algorithm. Similarly,~\cite{yu2023neuralecho} explored the joint training of AEC, SE, and AGC. However, their AGC targets were generated by a proprietary in-house AGC tool, which may cap the model's achievable performance and hinder reproducibility.

To address these limitations, this paper proposes SE-AGCNet, a deep-learning-based framework for joint SE and AGC. The proposed approach is designed for meeting-room scenarios, where audio quality is often degraded by background noise, reverberation, and volume discrepancies arising from differences in distance from the microphone, variations in speaking style, and head movements or shifts in body position. Experimental results demonstrate that SE-AGCNet achieves superior performance in both SE and AGC tasks, as well as in downstream Automatic Speech Recognition (ASR) tasks, achieving significant reductions in WER/CER while maintaining optimal loudness characteristics compared with conventional methods.

Our main contributions are summarized as follows:
\begin{itemize}
\item We propose SE-AGCNet\footnote{Code and demo: \texttt{https://jinming00.github.io/SE-AGCNet/}}, a flexible framework for joint, end-to-end training of SE and AGC. This framework supports integration with diverse existing speech enhancement architectures.
\item We propose SE-AGC-DataGen, a comprehensive and reproducible data simulation pipeline that addresses the lack of publicly available AGC datasets and enables joint SE-AGC training.
\item We introduce standardized loudness evaluation metrics (integrated loudness (LUFS), short-term loudness (St LUFS), and Loudness Range (LRA)) based on ITU-R BS.1770~\cite{itu1770} and EBU R128~\cite{ebuR128} to enable more objective and perceptually meaningful evaluation of AGC performance.

\end{itemize}

\begin{figure*}[t]
    \centering
    \includegraphics[width=0.98\textwidth]{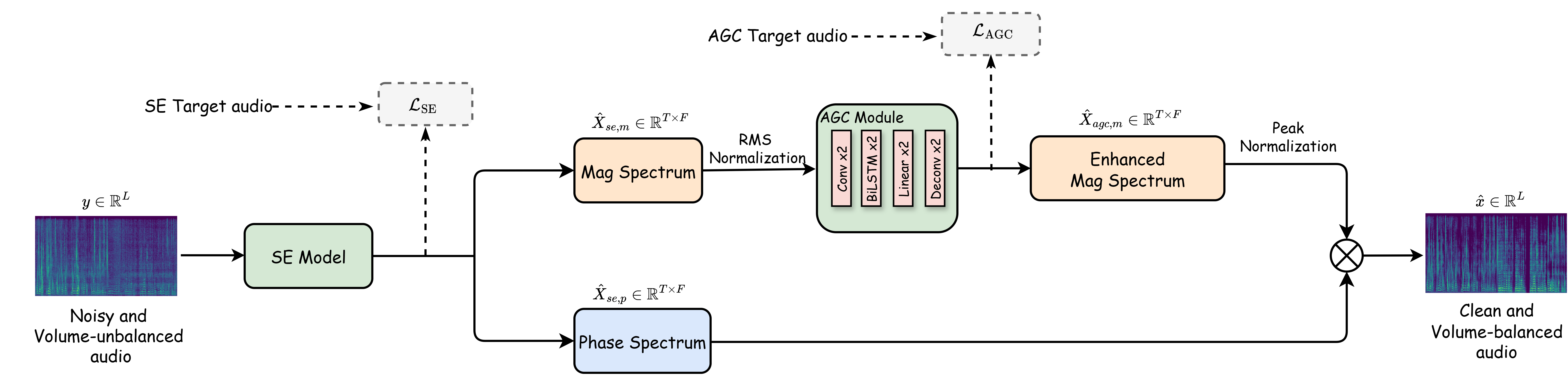}
    \caption{Overview of SE-AGCNet architecture. The system processes input audio through speech enhancement and automatic gain control modules in a joint training framework. SE Target Audio: clean and volume-unbalanced speech for SE module training; AGC Target Audio: clean and volume-balanced speech for AGC module training.}
    \label{fig:overview}
\end{figure*}

\section{Proposed SE-AGCNet}
\label{sec:method}
Figure~\ref{fig:overview} illustrates the SE-AGCNet architecture. Our approach employs a two-stage architecture that jointly optimizes speech enhancement and automatic gain control in the time-frequency domain, transforming a noisy and volume-unbalanced speech waveform $y \in \mathbb{R}^L$ into a clean and volume-balanced output $\hat{x} \in \mathbb{R}^L$, where $L$ denotes the waveform length.

Specifically, we first extract the magnitude spectrum $Y_m \in \mathbb{R}^{T \times F}$ and phase spectrum $Y_p \in \mathbb{R}^{T \times F}$ from the input waveform $y$ through STFT, where $T$ and $F$ denote the number of frames and frequency bins, respectively. The speech enhancement module processes the noisy input to produce enhanced magnitude spectrum $\hat{X}_{se,m} \in \mathbb{R}^{T \times F}$ and enhanced phase spectrum $\hat{X}_{se,p} \in \mathbb{R}^{T \times F}$:
\vspace{-2pt}
\begin{equation}
\hat{X}_{se,m}, \hat{X}_{se,p} = \text{SE}(Y_m, Y_p; \theta_{se})
\end{equation}
where $\theta_{se}$ denotes the speech enhancement model parameters. The enhanced magnitude spectrum is subsequently RMS-normalized and processed by the AGC module to generate a volume-balanced magnitude spectrum:
\vspace{-2pt}
\begin{equation}
\hat{X}_{agc,m} = \text{AGC}(\text{RMS-norm}(\hat{X}_{se,m}); \theta_{agc})
\end{equation}
where $\theta_{agc}$ denotes the AGC module parameters. Finally, the enhanced and volume-balanced waveform $\hat{x}$ is reconstructed via ISTFT using the AGC-processed magnitude spectrum and the enhanced phase from the SE module.

SE-AGCNet addresses two fundamental challenges: (1) the SE module performs noise suppression while preserving low-volume speech; and (2) the AGC module handles loudness adjustment. Through joint training, these modules operate synergistically in an end-to-end optimization framework.

\subsection{Speech Enhancement Model}
\label{ssec:SE model}
We adopt MP-SENet as the SE backbone and keep its original architecture and training setup (16 kHz sampling rate, 400-sample STFT window, 100-sample hop size, 400-point FFT, and 2-second training segments).

Our only modification is an asymmetric reweighting strategy~\cite{wang2020voicefilter}. Specifically, the MP-SENet loss definition remains unchanged, but for each time-frequency bin $(t,f)$, if the predicted magnitude is lower than the target magnitude (i.e., $\hat{X}_{se,m}^{t,f} < X_{se,target,m}^{t,f}$), we multiply the corresponding SE loss contribution by $\alpha=10.0$; otherwise, we keep its original weight. This imposes a $10\times$ stronger penalty on over-suppression.

During SE training, the target is clean but volume-unbalanced speech, so the model learns noise suppression while preserving loudness variation for the downstream AGC module.

\vspace{-3pt}
\subsection{AGC Module}
\label{ssec:agc module}

The AGC module performs volume adjustment by processing the RMS-normalized enhanced magnitude spectrum from MP-SENet to produce volume-balanced outputs. 
\vspace{-2pt}
\subsubsection{Architecture Design}

The AGC module takes the RMS-normalized enhanced magnitude spectrum $\hat{X}_{se,m}^{norm} \in \mathbb{R}^{T \times F}$ from the SE module as input and produces volume-balanced magnitude spectrum $\hat{X}_{agc,m} \in \mathbb{R}^{T \times F}$ as output. During training, the AGC target audio is the clean and volume-balanced magnitude spectrum $X_{agc,target,m}$.

The AGC module consists of three components: frequency-domain convolutional processing, bidirectional LSTM processing, and spectral reconstruction. The frequency-domain component extracts features through two 2D convolution layers:
\vspace{-4pt}
\begin{equation}
H_1 = \text{ReLU}(\text{BN}(\text{Conv2D}(\hat{X}_{se,m}^{norm})))
\end{equation}
\begin{equation}
H_{conv} = \text{ReLU}(\text{BN}(\text{Conv2D}(H_1)))
\end{equation}
where $H_{conv} \in \mathbb{R}^{T \times F \times 16}$ represents frequency-aware features extracted by 16-channel convolutions with $3 \times 3$ kernels. The bidirectional LSTM processes temporal sequences by reshaping the features to $(T, F \times 16)$ and using a 2-layer BiLSTM with hidden size 256 per direction. The LSTM output is then projected back to $F \times 16$ through linear layers before reconstruction. Finally, the reconstruction component employs transposed convolutions:
\vspace{-4pt}
\begin{equation}
\hat{X}_{agc,m} = \text{ReLU}(\text{ConvT}(\text{ReLU}(\text{BN}(\text{ConvT}(H_{lstm})))))
\end{equation}
with progressive channel reduction ensuring non-negative magnitude outputs.
\vspace{-5pt}
\subsubsection{Normalization Strategy}

The AGC module employs independent RMS normalization for both input and target during training, enabling the model to learn relative amplitude control rather than absolute values. The AGC processing includes peak normalization to 0.4 to achieve the target loudness of -23 LUFS. Since the AGC module precisely controls amplitude variations including sudden spikes, this peak normalization remains stable and avoids the common issue where peak normalization reduces legitimate speech to extremely low levels due to spike interference.
\vspace{-3pt}
\subsubsection{Training Objectives and Strategy}

To suppress noise amplification in silent target regions, we apply conditional weighting to the AGC loss, following the same reweighting principle used in the SE module:
\vspace{-1pt}
\begin{equation}
\resizebox{0.7\columnwidth}{!}{$\displaystyle
\mathcal{L}_{\mathrm{AGC}}=
\frac{1}{TF}\sum_{t=1}^{T}\sum_{f=1}^{F}
w_{t,f}\left|\hat{X}_{agc,m}^{t,f}-X_{agc,target,m}^{t,f}\right|
$}
\end{equation}
\begin{equation}
\resizebox{0.7\columnwidth}{!}{$\displaystyle
w_{t,f}=
\begin{cases}
10, & \text{if } X_{agc,target,m}^{t,f}=0 \land \hat{X}_{agc,m}^{t,f}>0 \\
1, & \text{otherwise}
\end{cases}
$}
\end{equation}
where $(t,f)$ denotes the time-frequency bin index, $\hat{X}_{agc,m}$ is the AGC output magnitude spectrum, and $X_{agc,target,m}$ is the target magnitude spectrum. This mechanism applies a $10\times$ stronger penalty when the AGC predicts positive energy in silent target regions, suppressing noise amplification.

The final training objective is defined as:
\vspace{-3pt}
\begin{equation}
\mathcal{L}_{\mathrm{total}} = \mathcal{L}_{\mathrm{MP-SENet}} + \lambda_{\mathrm{AGC}} \times \mathcal{L}_{\mathrm{AGC}}
\end{equation}
where $\mathcal{L}_{\mathrm{MP-SENet}}$ denotes the original MP-SENet multi-loss configuration with the asymmetric reweighting strategy described above, and $\lambda_{\mathrm{AGC}} = 0.9$, which is consistent with the magnitude-loss weight in MP-SENet.

To ensure stability, we adopt a curriculum learning strategy: the MP-SENet module is pre-trained for 5 epochs using SE targets before jointly optimizing the entire framework.

\section{SE-AGC-DataGen Data Simulation Pipeline}
\label{sec:dataset}

Since no publicly available dataset exists for AGC tasks, we develop a comprehensive data simulation pipeline that generates multi-speaker audio with realistic volume variations and acoustic conditions. We create one simulated dataset, LibriAGC, for training and evaluation.
LibriTTS~\cite{zen19_interspeech} is used as our base data source, as it contains relatively volume-balanced clean speech suitable for controlled volume variation simulation. 

Our simulation process consists of three stages: \textbf{Stage 1:} We concatenate clean speech segments from 2-5 speakers. \textbf{Stage 2:} We apply a two-layer volume processing mechanism that includes basic volume adjustment (100\% original or reduced to 5\%-30\%) and four audio augmentation modes (sudden spikes, gradual increase/decrease, and volume fluctuations, each applied with 15\% probability). These augmentation modes simulate real-world scenarios where volume variations occur due to speaker emotions, changing distances from microphones, recording equipment characteristics, and audio clipping artifacts encountered in meeting scenarios. \textbf{Stage 3:} We specially select 35 noise clips from the DNS Challenge noise set and mix them at 5--25 dB SNR to mimic meeting-room acoustics, including common noises such as fan hum, keyboard typing, door sounds, and chair movement. 

For training targets, we generate two reference signals from different processing stages: the \textbf{SE target} is clean but volume-unbalanced speech from Stage 2, used to train the speech enhancement module; the \textbf{AGC target} is clean and volume-balanced speech from Stage 1, used to train the AGC module. In Table~\ref{tab:comprehensive_performance}, \textit{Ref} denotes this AGC target, while \textit{Input} denotes the simulated noisy and volume-unbalanced audio after Stage 3.

Following the SE-AGC-DataGen pipeline, we construct one simulated dataset for training and evaluation: \textbf{LibriAGC}, built from LibriTTS-train-clean-100 (train) and LibriTTS-test-clean (test). The final dataset contains 9,487 training utterances (54 hours) and 1,406 test utterances (8 hours). The effectiveness of this simulation pipeline is further validated in the real-world evaluation results in Section~\ref{ssec:realworld}.

\begin{table}[h]
    \caption{Loudness statistics of high-quality clean speech datasets and raw real-world recordings. For VoiceBank+DEMAND and LibriTTS, we concatenate 2--5 clean utterances to form clips of at least 10 s and report LUFS, St LUFS, and LRA (LU).}
    \label{tab:loudness_stats}
    \centering
    \resizebox{0.9\columnwidth}{!}{
    \begin{tabular}{l ccc}
        \toprule
        \textbf{Dataset} & \textbf{LUFS} & \textbf{St LUFS} & \textbf{LRA} \\
        \midrule
        VoiceBank+DEMAND & -23.38 & -25.06 & 3.68 \\
        LibriTTS & -23.95 & -24.77 & 4.33 \\
        \hdashline
        MMCSG  & -40.24 & -45.92 & 20.21 \\
        AliMeeting-far  & -34.89 & -37.40 & 12.08 \\
        \bottomrule
    \end{tabular}
    }
\end{table}

\begin{table*}[!htbp] 
    \caption{Evaluation on LibriAGC test set: LUFS/St LUFS, LRA (LU), WER (\%). AGC metrics meeting the target requirements (LUFS and St LUFS around -23, LRA between 3-6 LU) are \underline{underlined}, while the best results for other metrics are \textbf{bolded}. Ref: Clean and volume-balanced reference; Input: Noisy and volume-unbalanced input. Models marked with $\dagger$ are re-trained on LibriAGC dataset. $^a$Whisper-large-v3-turbo~\cite{radford2023robust}; $^b$nvidia/stt\_en\_conformer\_ctc\_large~\cite{kuchaiev2019nemo}.}
    \label{tab:comprehensive_performance}
    \vspace{2pt} 
    \centering
    
    \resizebox{1.0\textwidth}{!}{
    \begin{tabular}{l c cccc ccc ccccc}
        \toprule
        \multirow{2}{*}{\textbf{System}} & 
        \multirow{2}{*}{PESQ $\uparrow$} &
        \multicolumn{4}{c}{SIGMOS $\uparrow$} &
        \multicolumn{3}{c}{DNSMOS $\uparrow$} &
        \multirow{2}{*}{LUFS} &
        \multirow{2}{*}{St LUFS} &
        \multirow{2}{*}{LRA} &
        \multirow{2}{*}{WER$^a$ $\downarrow$} &
        \multirow{2}{*}{WER$^b$ $\downarrow$} \\
        
        \cmidrule(lr){3-6} \cmidrule(lr){7-9}
        
        & & SIG & NOISE & LOUD & OVRL & SIG & BAK & OVRL & & & & & \\
        \midrule
        
        \textit{Ref} & - & 3.53 & 3.78 & 3.91 & 3.13 & 3.64 & 4.08 & 3.36 & -22.75 & -23.36 & 4.21 & 2.40 & 3.69 \\
        \textit{Input} & 1.33 & 2.87 & 2.55 & 3.18 & 2.41 & 3.34 & 2.89 & 2.57 & -24.12 & -28.95 & 14.87 & 10.67 & 21.61 \\
        \hdashline

        \textit{MP-SENet (Orig)} & 1.55 & 3.36 & \textbf{4.00} & 3.50 & 2.98 & 3.44 & 3.93 & 3.10 & -23.90 & -30.61 & 18.69 & 21.26 & 27.16 \\
        \quad + \textit{pyagc} & 1.63 & 3.18 & 3.61 & 3.50 & 2.84 & 3.49 & 3.77 & 3.10 & \underline{-23.03} & \underline{-23.29} & \underline{3.77} & 20.09 & 23.15 \\

        \textit{MP-SENet (SE)}$^\dagger$ & 2.18 & \textbf{3.38} & 3.99 & 3.47 & 2.98 & 3.52 & 4.10 & 3.24 & -23.76 & -30.52 & 18.61 & 7.61 & 11.20 \\
        \quad + \textit{pyagc} & 2.69 & 3.25 & 3.58 & 3.47 & 2.89 & 3.56 & 4.05 & 3.29 & -20.88 & -21.10 & \underline{3.49} & 7.09 & 9.35 \\

        \textit{MP-SENet (AGC)}$^\dagger$ & 2.80 & 3.35 & 3.68 & 3.86 & 2.94 & 3.55 & 4.09 & 3.29 & -24.74 & -26.19 & 8.32 & 7.69 & 10.61 \\
                      
        \textit{SE-AGCNet}$^\dagger$ & \textbf{3.00} & \textbf{3.38} & 3.68 & \textbf{3.87} & \textbf{2.99} & \textbf{3.60} & \textbf{4.11} & \textbf{3.35} & \underline{-23.66} & \underline{-23.93} & \underline{3.86} & \textbf{6.88} & \textbf{9.12} \\

        \bottomrule
    \end{tabular}
    }
\end{table*}

\begin{table}[!htbp] 
    \caption{Evaluation on real-world datasets: LUFS/St LUFS, LRA (LU), WER (\%), CER (\%). Models marked with $\dagger$ are re-trained on LibriAGC dataset.}
    \label{tab:performance_datasets}
    \vspace{2pt}
    \centering
    
    \resizebox{\columnwidth}{!}{
    \begin{tabular}{l ccccc cccc}
        \toprule
        \multirow{2}{*}{\textbf{System}} & 
        \multicolumn{5}{c}{\textbf{MMCSG}} & 
        \multicolumn{4}{c}{\textbf{AliMeeting-far}} \\
        \cmidrule(lr){2-6} \cmidrule(lr){7-10}
        & LUFS & St LUFS & LRA & WER$^a$ $\downarrow$ & WER$^b$ $\downarrow$ & LUFS & St LUFS & LRA & CER$^a$ $\downarrow$ \\
        \midrule
        \textit{Noisy} & -40.24 & -45.92 & 20.21 & 15.12 & 51.36 & -34.89 & -37.40 & 12.08 & 36.16 \\
        \hdashline
        
        \textit{MP-SENet (Orig)} & -38.86 & -45.23 & 23.15 & 46.74 & 52.80 & -40.35 & -47.12 & 22.67 & 81.91 \\
        \quad + \textit{pyagc} & -29.91 & -30.42 & \underline{5.72} & 44.25 & 50.72 & -34.78 & -35.41 & 6.38 & 79.98 \\

        \textit{MP-SENet (SE)}$^\dagger$ & -39.07 & -45.16 & 20.83 & 14.41 & 50.85 & -35.66 & -38.25 & 12.36 & 38.95 \\
        \quad + \textit{pyagc} & \underline{-23.74} & \underline{-24.08} & \underline{4.65} & 14.06 & 34.91 & -28.98 & -29.13 & 2.83 & 36.78 \\

        \textit{MP-SENet (AGC)}$^\dagger$ & -47.57 & -53.08 & 19.72 & 15.19 & 51.92 & -32.58 & -34.66 & 11.66 & 43.06 \\
                      
        \textit{SE-AGCNet}$^\dagger$ & \underline{-22.68} & \underline{-23.04} & \underline{4.80} & \textbf{13.86} & \textbf{30.36} & \underline{-22.89} & \underline{-23.20} & \underline{4.26} & \textbf{34.43} \\
        \bottomrule 
    \end{tabular}
    }
\end{table}

\vspace{-6pt}
\section{Experimental Setup and Results}
\label{sec:experiments}
\subsection{Datasets}
\label{ssec:exp_datasets}
\vspace{-4pt}
We use the following datasets in our experiments: \textbf{VoiceBank+DEMAND}~\cite{botinhao2016investigating}, which is used in Section~\ref{ssec:metrics} as a clean-speech loudness reference (Table~\ref{tab:loudness_stats}); \textbf{LibriAGC}, a simulated dataset described in Section~\ref{sec:dataset}; and two real-world datasets, \textbf{MMCSG}~\cite{zmolikova2024chime}, a CHiME-8 challenge dataset with two-person conversation recordings using Aria glasses (evaluation set: 189 files, 9.4 hours), and \textbf{AliMeeting-far}~\cite{Yu2022M2MeT}, the far-field speech portion from AliMeeting test set (20 files, 10.8 hours). For multi-channel audio, we use the first channel for evaluation. Both real-world datasets exhibit significant volume imbalance due to varying speaker distances and acoustic conditions.
\vspace{-6pt}
\subsection{Evaluation Metrics}
\label{ssec:metrics}
\vspace{-3pt}

\textbf{Subjective metrics:} We use SIGMOS~\cite{ristea2025icassp} and DNSMOS~\cite{reddy2021dnsmos} for perceptual quality assessment. SIGMOS includes loudness MOS, making it particularly well-suited for AGC evaluation.

\textbf{Objective metrics:} Speech quality is also assessed using PESQ. For speech recognition performance, we compute WER and CER to evaluate the impact on downstream applications.

\textbf{AGC-specific metrics:} We introduce standardized loudness metrics based on ITU-R BS.1770 standards and EBU R128 recommendations to evaluate AGC performance. These metrics address limitations of conventional approaches like absolute gain and RMS that may not accurately reflect AGC effectiveness.

\textit{LUFS (Loudness Units relative to Full Scale):} LUFS provides a perceptually-weighted loudness measure that better correlates with human auditory perception than traditional RMS measurements. We set the target loudness to -23 LUFS by following the EBU R128 loudness normalization recommendation. We then analyze the loudness statistics of two high-quality clean speech datasets using VoiceBank+DEMAND train-clean and LibriTTS test-clean to verify this setting. Table~\ref{tab:loudness_stats} shows that their mean LUFS values are close to -23 LUFS with moderate LRA, supporting -23 LUFS as a reasonable loudness target for clear speech. In contrast, raw real-world recordings (MMCSG and AliMeeting-far) exhibit substantially lower loudness and much larger LRA, motivating the need for AGC; SE-AGCNet restores their loudness to the target range as reported in Table~\ref{tab:performance_datasets}.

\textit{St LUFS (Short-term LUFS):} Standard LUFS employs gating mechanisms that exclude relatively very quiet audio segments, which can produce biased measurements in meeting room scenarios where near-field and far-field speech should be treated equally. For example, in Table~\ref{tab:comprehensive_performance}, despite significant volume variations being applied, the LUFS difference between \textit{Ref} and \textit{Input} is modest due to the gating effect. To address this limitation, we utilize Short-term LUFS, defined as loudness measurements using a 3-second sliding window with a 1-second hop, averaged across the entire audio segment. This approach reduces gating influence and provides more physically accurate loudness estimation. For effective AGC evaluation, we require both LUFS and St LUFS to achieve values around -23.

\textit{LRA (Loudness Range):} LRA quantifies the dynamic range of an audio signal based on short-term LUFS. Extremely high LRA indicates severe volume imbalance, while excessively low LRA may result in over-compressed audio with reduced dynamics. Based on our reference audio analysis, we consider LRA values between 3-6 LU as optimal for volume-balanced speech with appropriate dynamic range preservation.
\vspace{-6pt}
\subsection{Baseline Systems}
\label{ssec:baselines}
\vspace{-3pt}
We compare SE-AGCNet against several baselines:

\textbf{MP-SENet (Orig):} The original MP-SENet checkpoint, downloaded from the official GitHub repository, which was trained on the DNS Challenge 2020 dataset. 

\textbf{MP-SENet (SE):} MP-SENet retrained with noisy and volume-unbalanced audio as input and clean and volume-unbalanced audio as target (i.e., SE target audio). This configuration focuses solely on speech enhancement without AGC functionality.

\textbf{MP-SENet (AGC):} MP-SENet retrained with noisy and volume-unbalanced audio as input but targeting clean and volume-balanced audio (i.e., AGC target audio), attempting to learn both speech enhancement and AGC simultaneously within a single model.

\textbf{+ \textit{pyagc}:} We use \textit{pyagc}\footnote{https://github.com/jorgehatccrma/pyagc} as a post-processing baseline for conventional AGC behavior. It is the most-forked open-source AGC implementation on GitHub, adapted from the MATLAB implementation by Ellis~\cite{ellis2026agc}, and follows an attack/release-time-based control strategy. To the best of our knowledge, few newer open-source AGC tools are publicly available. This baseline enables a fair comparison between our joint learning approach and an off-the-shelf AGC post-processor.

For both MP-SENet (SE) and MP-SENet (AGC) above, we use LibriAGC training data for experiments in Sections~\ref{ssec:simulated}. At inference time, long recordings are processed using 2-second windows with 50\% overlap.


\vspace{-4pt}
\subsection{Evaluation on LibriAGC}

\label{ssec:simulated}

Table~\ref{tab:comprehensive_performance} presents evaluation results on the LibriAGC test set. The open-source MP-SENet exhibits severe over-suppression of far-field speech, resulting in poor PESQ and high WER. A key reason is that the released checkpoint was not trained on data with such large loudness variation, since publicly available datasets of this kind were limited before. In particular, MP-SENet (Orig) + \textit{pyagc} corresponds to the conventional cascaded processing strategy (SE followed by AGC). Compared with MP-SENet (SE), SE-AGCNet improves SE quality, ASR performance, and loudness control, showing clear gains from adding the AGC module. Compared with MP-SENet (SE) + \textit{pyagc}, SE-AGCNet still shows consistent gains on the reported metrics, suggesting that joint optimization is more effective than using AGC only as a separate post-processing step.

\vspace{-6pt}
\subsection{Evaluation on Real-world Datasets}
\label{ssec:realworld}

Table~\ref{tab:performance_datasets} validates SE-AGCNet's applicability on real-world datasets. To assess practical impact on downstream recognition, we use recognition error rates as key metrics (WER for MMCSG and CER for AliMeeting-far). WER/CER are jointly affected by residual noise, speech distortion, and loudness inconsistency, and therefore reflect both SE effectiveness and AGC quality in practical use. SE-AGCNet consistently achieves target AGC characteristics across both datasets while maintaining the best ASR performance. Notably, the relative improvements are more pronounced with the ASR model trained on limited data (WER$^b$) compared to the highly robust Whisper model (WER$^a$), which has been trained on extensive and diverse speech scenarios. This suggests that our approach provides greater benefits for ASR systems with constrained training data that are more sensitive to audio quality variations.

\vspace{-5pt}
\section{Conclusion}
\label{sec:conclusion}

This paper presents SE-AGCNet, a joint framework for speech enhancement (SE) and automatic gain control (AGC) that addresses the limitations of cascaded pipelines. Through end-to-end optimization, SE and AGC are trained to cooperate, allowing SE to preserve low-volume speech while AGC adjusts loudness. We also introduce SE-AGC-DataGen for SE-AGC tasks and adopt standardized loudness metrics (LUFS, St LUFS, and LRA). Experiments on simulated and real-world meeting datasets show that SE-AGCNet jointly improves speech enhancement and loudness control: it suppresses noise while preserving speech content, brings loudness close to target ranges, and improves downstream ASR performance. The modular design also supports integration with existing SE backbones. Future work will extend the framework to more complex acoustic conditions and explore lightweight real-time models.

\section{Generative AI Use Disclosure}
Generative AI tools were used only for language editing and polishing. They were not used to generate a substantial portion of the manuscript. All AI-assisted edits were carefully reviewed and revised by the authors, who take full responsibility for the paper and approve its submission.

\bibliographystyle{IEEEtran}
\bibliography{mybib}

\end{document}